\documentclass[12pt]{article}%
\usepackage{amsfonts}
\usepackage{graphics}
\usepackage{amsmath}
\usepackage{amssymb}
\usepackage{graphicx}%
\providecommand{\U}[1]{\protect\rule{.1in}{.1in}}
\begin{document}
\centerline{\textbf{\Large}}

\vskip 0.8truecm
\centerline{\textbf{\Large Relativistic corrections to the Zeeman splitting}}
\vskip 0.3truecm \centerline{\textbf{\Large of hyperfine structure levels}}
\vskip 0.3truecm
\centerline{\textbf{\Large in two-fermion bound-state systems}}
\vskip 0.6truecm \centerline{\large Andrei G. Terekidi$^{a }$,
Jurij W. Darewych$^{b }$, Marko Horbatsch$^{c}$}

\vskip 0.5truecm \centerline{\footnotesize \emph{Department of Physics and
Astronomy, York University, Toronto, Ontario, M3J 1P3, Canada}}
\centerline{\footnotesize \emph{$^{a }$terekidi@yorku.ca,
$^{b}$darewych@yorku.ca,$^{c}$marko@yorku.ca } }

\vskip1.6truecm

\begin{center}
\noindent\textbf{Abstract}
\end{center}

A relativistic theory of the Zeeman splitting of hyperfine levels in
two-fermion systems is presented. The approach is based on the variational
equation for bound states derived from quantum electrodynamics [1].
Relativistic corrections to the $g$-factor are obtained up to $O\left(
\left(  \alpha\right)  ^{2}\right)  $. Calculations are provided for all
quantum states and for arbitrary fermionic mass ratio. In the one-body limit
our calculations reproduce the formula for the $g$-factor (to $O\left(
\left(  Z\alpha\right)  ^{2}\right)  $) obtained from the Dirac equation. The
results will be useful for comparison with high-precision measurements.

\vskip0.6truecm

\noindent\textbf{{\large 1. Introduction}}

\vskip0.4truecm

In a recent paper [2] we have presented a self-consistent variational method
for calculating the non-relativistic Land\'{e} $g$ factor of the two-fermion
bound-state system. In the lowest-order approximation the linearly dependent
part of the energy splitting for a two-fermion system placed in a weak static
magnetic field $\mathbf{B}$\ can be written as [2-6]%
\begin{equation}
\Delta E_{J,m_{J},S,\ell,s_{1},s_{2}}^{ext}=\left(  \mu_{B1}g_{1}+\mu
_{B2}g_{2}\right)  Bm_{J},
\end{equation}
where $J$, $m_{J}$, $S$, $\ell$, $s_{1}$, $s_{2}$\ are quantum numbers, which
characterize the system: $s_{1}$ and $s_{2}$ are the spins of the first and
second particle respectively, $S=s_{1}+s_{2},\ s_{1}+s_{2}-1,...,\ \left\vert
s_{1}-s_{2}\right\vert $\ is the total spin of the particles, $\ell$ and
$J$\ represent the orbital and total angular momentum quantum numbers, where
$J=\ell+S,$\ $\ell+S-1,...,\ \left\vert \ell-S\right\vert $. The projection of
the total angular momentum on the $\mathbf{B}$ field direction is labeled by
$m_{J}=-J,\ -J+1,...,\ J-1,\ J$. The \textquotedblleft Bohr
magnetons\textquotedblright\ for the two particles are defined as
$\displaystyle\mu_{B1}=Q_{1}\hbar/2m_{1}c$, and $\mu_{B2}=-Q_{2}\hbar/2m_{2}%
c$, where $Q_{1}$ and $Q_{2}$ represent the magnitude of the charges. In our
notation, $m_{1}$\ and $m_{2}$\ correspond to the masses of the light and
heavy particle respectively. The description of the interacting system by the
set of quantum numbers $J$, $m_{J}$, $S$, $\ell$, $s_{1}$, $s_{2}$ corresponds
to the $LS$\ coupling representation. This representation is used in contrast
to the customary $j_{1}$-$j_{2}$\ coupling scheme (for the case $m_{2}>>m_{1}%
$), where the states are taken to be the eigenstates of the operators
$\widehat{\mathbf{j}}_{1}^{2}=\left(  \widehat{\mathbf{L}}+\widehat
{\mathbf{s}}_{1}\right)  ^{2}$, $\widehat{\mathbf{j}}_{2}^{2}=\widehat
{\mathbf{s}}_{2}^{2}$. As discussed previously [7], for the general case of
arbitrary mass ratio, the $j_{1}$\ value is not a good quantum number. Even in
the $LS$\ representation, for spin-mixed states the orbital angular momentum
$L$, and total spin $S$ of the system are not conserved. In this case we
designate the states by an additional quantum number $\widetilde{s}$, which
takes on the values of $0$ or $1$ for quasisinglet ($sg_{q}$) and quasitriplet
($tr_{q}$) states respectively.

In our calculations we assume that the energy-level splitting (1) is smaller
than the hyperfine structure (HFS) splitting, $\Delta E^{ext}<<\Delta E^{HFS}%
$, i.e., we treat the interaction with an external magnetic field as a perturbation.

The non-relativistic Land\'{e} factors $g_{1}$\ and $g_{2}$ obtained in [2]
can be summarized as follows:

\vskip0.4truecm

\noindent for $\ell=J-1$:%
\begin{equation}
g_{i}=1-\mu_{i}\frac{J-1}{J}+\left(  \frac{\widetilde{g}_{s_{i}}}{2}-1\right)
\frac{1}{J},
\end{equation}
\noindent for $\ell=J+1$:%
\begin{equation}
g_{i}=1-\mu_{i}\frac{J+2}{J+1}-\left(  \frac{\widetilde{g}_{s_{i}}}%
{2}-1\right)  \frac{1}{J+1},
\end{equation}
for \textit{spin--mixed} states $\ell=J\neq0$%
\begin{equation}
g_{i}=\left(  1-\frac{1\mp\xi}{2J\left(  J+1\right)  }\right)  \mu_{j}%
+\frac{\widetilde{g}_{s_{i}}}{2}\left(  \frac{1\mp\xi}{2J\left(  J+1\right)
}\mp\left(  -1\right)  ^{i}2\left\vert \mu_{i}-\mu_{j}\right\vert \xi\right)
,
\end{equation}
where $i=1,2$ is the index designating the particle. The index $j$ is defined
as $j=1$ when $i=2$, and $j=2$ when $i=1$. The quantities $\mu_{i}$ represent
the mass factors:%
\begin{equation}
\mu_{i}=\frac{m_{i}}{m_{1}+m_{2}},
\end{equation}
and $\xi$\ is given as%
\begin{equation}
\xi=\left(  4\left(  \mu_{1}-\mu_{2}\right)  ^{2}J\left(  J+J\right)
+1\right)  ^{-1/2}.
\end{equation}
The upper and the lower signs in (4) correspond to the quasisinglet $sg_{q}$
and quasitriplet $tr_{q}$ states respectively. Note that $\widetilde{g}%
_{s_{i}}$\ are the intrinsic spin magnetic moments of the constituent
particles. According to the Dirac theory a free particle at rest has
$\widetilde{g}_{s_{i}}=2$. In QED $\widetilde{g}_{s_{i}}$ is modified by the
\textquotedblleft anomaly\textquotedblright, which to lowest order is given by
the Schwinger correction $\widetilde{g}_{s_{i}}=2+\alpha/\pi$ [7], where
$\alpha$\ is the fine-structure constant.

In the case when $m_{2}>>m_{1}$\ our general result reduces to the previously
known result [3-6], in which the orbital motion of the heavy particle is
ignored, namely%
\begin{equation}
g_{1}=g_{j_{1}}\frac{F\left(  F+1\right)  +j_{1}\left(  j_{1}+1\right)
-I\left(  I+1\right)  }{2F\left(  F+1\right)  },
\end{equation}
where%
\begin{equation}
g_{j_{1}}=1+\left(  \widetilde{g}_{s_{1}}-1\right)  \frac{j_{1}\left(
j_{1}+1\right)  +s_{1}\left(  s_{1}+1\right)  -\ell\left(  \ell+1\right)
}{2j_{1}\left(  j_{1}+1\right)  },
\end{equation}
and%
\begin{equation}
g_{2}=\widetilde{g}_{s_{2}}\frac{F\left(  F+1\right)  -j_{1}\left(
j_{1}+1\right)  +I\left(  I+1\right)  }{2F\left(  F+1\right)  }.
\end{equation}
Here $I$\ is the spin of the second particle, and $F$\ is the total angular
momentum of the system. To facilitate the comparison of (4) with (7)-(9) we
need to make the following replacement of quantum numbers: $F\rightarrow J$,
$J\rightarrow j_{1}$, $\ell_{1}=\ell$, $S\rightarrow s_{1}$, $I\rightarrow
s_{2}$.

In this paper we present results for the relativistic case beyond the formulae
(2)-(4). Thus, the $g$-factors in (1) are written more generally as
\begin{equation}
g_{i}=g_{i}^{NR}+\bigtriangleup g_{i}^{REL},
\end{equation}
where $g_{i}^{NR}$\ are the nonrelativistic Land\'{e} factors defined by
(2)-(4), and $\bigtriangleup g_{i}^{REL}$ is the relativistic correction. In
the next section we calculate $\bigtriangleup g_{i}^{REL}$ to order $O\left(
\left(  \alpha^{\prime}\right)  ^{2}\right)  $\ for all quantum states and
arbitrary masses of particles. In most expressions we use natural units, i.e.,
$\hbar=c=1$, $\alpha=e^{2}/4\pi$. The coupling constant is defined as
$\alpha^{\prime}=Q_{1}Q_{2}/4\pi$.

\vskip0.4truecm

\noindent\textbf{{\large 2. Variational wave equation and relativistic
corrections to g-factors of two-fermion systems}}

\vskip0.4truecm

The relativistic wave equations for two-fermion systems in the absence of
external fields were derived in [1] and [8] on the basis of a modified QED
Lagrangian [9]-[10]. In this approach a simple Fock-space trial state of the
form%
\begin{equation}
\left\vert \psi_{trial}\right\rangle =\underset{s_{1}s_{2}}{\sum}\int
d^{3}\mathbf{p}_{1}d^{3}\mathbf{p}_{2}F_{s_{1}s_{2}}(\mathbf{p}_{1}%
,\mathbf{p}_{2})b_{\mathbf{p}_{1}s_{1}}^{\dagger}D_{\mathbf{p}_{2}s_{2}%
}^{\dagger}\left\vert 0\right\rangle ,
\end{equation}
is sufficient to obtain the HFS levels correct to fourth order in the coupling
constant $\alpha^{\prime}$. Here $b_{\mathbf{q}_{1}s_{1}}^{\dagger}$ and
$D_{\mathbf{q}_{2}s_{2}}^{\dagger}$\ are creation operators for a free fermion
of mass $m_{1}$ and an (anti) fermion of mass $m_{2}$ respectively, and
$\left\vert 0\right\rangle $ is the trial vacuum state such that
$b_{\mathbf{q}_{1}s_{1}}\left\vert 0\right\rangle =D_{\mathbf{q}_{2}s_{2}%
}\left\vert 0\right\rangle =0$. The $F_{s_{1}s_{2}}$ are four-component
adjustable functions.

The variational principle $\delta\left\langle \psi_{trial}\left\vert
\widehat{H}-E\right\vert \psi_{trial}\right\rangle =0$, where $\widehat{H}%
$\ is the QED Hamiltonian is invoked to obtain a momentum-space wave equation
for the amplitudes $F_{s_{1}s_{2}}$\ [1]:
\begin{align}
0  &  =\sum_{s_{1}s_{2}}\int d^{3}\mathbf{p}_{1}d^{3}\mathbf{p}_{2}\left(
\omega_{p_{1}}+\Omega_{p_{2}}-E\right)  F_{s_{1}s_{2}}(\mathbf{p}%
_{1},\mathbf{p}_{2})\delta F_{s_{1}s_{2}}^{\ast}(\mathbf{p}_{1},\mathbf{p}%
_{2})\\
&  -\frac{m_{1}m_{2}}{\left(  2\pi\right)  ^{3}}\underset{\sigma_{1}\sigma
_{2}s_{1}s_{2}}{\sum}\int\frac{d^{3}\mathbf{p}_{1}d^{3}\mathbf{p}_{2}%
d^{3}\mathbf{q}_{1}d^{3}\mathbf{q}_{2}}{\sqrt{\omega_{p_{1}}\omega_{q_{1}%
}\Omega_{p_{2}}\Omega_{q_{2}}}}\nonumber\\
&  \times F_{\sigma_{1}\sigma_{2}}(\mathbf{q}_{1},\mathbf{q}_{2})\left(
-i\right)  \widetilde{\mathcal{M}}_{s_{1}s_{2}\sigma_{1}\sigma_{2}}\left(
\mathbf{p}_{1},\mathbf{p}_{2},\mathbf{q}_{1}\mathbf{,q}_{2}\right)  \delta
F_{s_{1}s_{2}}^{\ast}(\mathbf{p}_{1},\mathbf{p}_{2}),\nonumber
\end{align}
where $\omega_{p_{1}}^{2}=\mathbf{p}_{1}^{2}+m_{1}^{2}$ and $\Omega_{p_{1}%
}^{2}=\mathbf{p}_{1}^{2}+m_{2}^{2}$. The inter-particle interaction is
represented by the generalized invariant $\mathcal{M}$-matrix,%
\begin{equation}
\widetilde{\mathcal{M}}_{s_{1}s_{2}\sigma_{1}\sigma_{2}}\left(  \mathbf{p}%
_{1},\mathbf{p}_{2},\mathbf{q}_{1}\mathbf{,q}_{2}\right)  =\mathcal{M}%
_{s_{1}s_{2}\sigma_{1}\sigma_{2}}^{\left(  1\right)  }\left(  \mathbf{p}%
_{1},\mathbf{p}_{2},\mathbf{q}_{1}\mathbf{,q}_{2}\right)  +\mathcal{M}%
_{s_{1}s_{2}\sigma_{1}\sigma_{2}}^{\left(  2\right)  }\left(  \mathbf{p}%
_{1},\mathbf{p}_{2},\mathbf{q}_{1}\mathbf{,q}_{2}\right)  +..,
\end{equation}
obtained as part of the derivation. It includes reducible and irreducible
effects in all orders of the coupling constant $\alpha^{\prime}$, and the sum
contains all relevant Feynman diagrams. A discussion of the derivation and
structure of the $\widetilde{\mathcal{M}}$-matrix\ to one-loop level is
provided in [1]. This equation allows one to obtain, in principle, all
relativistic and QED corrections to the $g$-factor.

The lowest-order QED corrections appear within the term $\mathcal{M}^{\left(
2\right)  }$ and can be formally included in the intrinsic factor
$\widetilde{g}_{s_{i}}$, however we shall not do so in this work. In this
paper we restrict our consideration to the first term $\mathcal{M}^{\left(
1\right)  }$\ of the expansion (13), i.e., only tree level diagrams are
included. The term $\mathcal{M}^{\left(  1\right)  }$\ contains only
relativistic corrections and it\ can be broken into two parts, namely%
\begin{equation}
\mathcal{M}_{s_{1}s_{2}\sigma_{1}\sigma_{2}}^{\left(  1\right)  }\left(
\mathbf{p}_{1},\mathbf{p}_{2},\mathbf{q}_{1}\mathbf{,q}_{2}\right)
=\mathcal{M}_{s_{1}s_{2}\sigma_{1}\sigma_{2}}^{ope}\left(  \mathbf{p}%
_{1},\mathbf{p}_{2},\mathbf{q}_{1}\mathbf{,q}_{2}\right)  +\mathcal{M}%
_{s_{1}s_{2}\sigma_{1}\sigma_{2}}^{ext}\left(  \mathbf{p}_{1},\mathbf{p}%
_{2},\mathbf{q}_{1}\mathbf{,q}_{2}\right)  ,
\end{equation}
where $\mathcal{M}_{s_{1}s_{2}\sigma_{1}\sigma_{2}}^{ope}\left(
\mathbf{p}_{1},\mathbf{p}_{2},\mathbf{q}_{1}\mathbf{,q}_{2}\right)  $ is the
usual invariant matrix element, corresponding to the one-photon exchange
Feynman diagram [1]. The element $\mathcal{M}_{s_{1}s_{2}\sigma_{1}\sigma_{2}%
}^{ext}$ represents the interaction with a given external classical field
$A_{\mu}^{ext}$,%
\begin{align}
&  \mathcal{M}_{s_{1}s_{2}\sigma_{1}\sigma_{2}}^{ext}\left(  \mathbf{p}%
_{1},\mathbf{p}_{2},\mathbf{q}_{1}\mathbf{,q}_{2}\right) \\
&  =i\left(  2\pi\right)  ^{3/2}\left(
\begin{array}
[c]{c}%
\frac{\sqrt{\Omega_{\mathbf{p}_{2}}\Omega_{q_{2}}}}{m_{2}}A_{\mu}%
^{ext}(\mathbf{p}_{1}-\mathbf{q}_{1})\overline{u}\left(  \mathbf{p}_{1}%
,s_{1}\right)  \left(  -iQ_{1}\right)  \gamma^{\mu}u\left(  \mathbf{q}%
_{1},\sigma_{1}\right)  \delta_{s_{2}\sigma_{2}}\\
+\frac{\sqrt{\omega_{\mathbf{p}_{1}}\omega_{\mathbf{q}_{1}}}}{m_{1}}A_{\mu
}^{ext}(\mathbf{q}_{2}-\mathbf{p}_{2})\overline{V}\left(  \mathbf{p}%
_{2},\sigma_{2}\right)  \left(  -iQ_{2}\right)  \gamma^{\mu}V\left(
\mathbf{q}_{2},s_{2}\right)  \delta_{s_{1}\sigma_{1}}%
\end{array}
\right)  .\nonumber
\end{align}
Using the semi-relativistic expansion of the expression $\overline{u}\left(
\mathbf{p}_{1},s_{1}\right)  \mathbf{\gamma}u\left(  \mathbf{q}_{1},\sigma
_{1}\right)  $ up to order $1/c^{3}$ we obtain for the $\mathcal{M-}$matrix%
\begin{equation}
\mathcal{M}_{s_{1}s_{2}\sigma_{1}\sigma_{2}}^{ext}\left(  \mathbf{p}%
_{1},\mathbf{p}_{2},\mathbf{q}_{1}\mathbf{,q}_{2}\right)  =\mathcal{M}%
_{s_{1}s_{2}\sigma_{1}\sigma_{2}}^{\left(  1\right)  ext}\left(
\mathbf{p}_{1},\mathbf{p}_{2},\mathbf{q}_{1}\mathbf{,q}_{2}\right)
+\mathcal{M}_{s_{1}s_{2}\sigma_{1}\sigma_{2}}^{\left(  2\right)  ext}\left(
\mathbf{p}_{1},\mathbf{p}_{2},\mathbf{q}_{1}\mathbf{,q}_{2}\right)  ,
\end{equation}
where%
\begin{align}
&  \mathcal{M}_{s_{1}s_{2}\sigma_{1}\sigma_{2}}^{\left(  1\right)  ext}\left(
\mathbf{p}_{1},\mathbf{p}_{2},\mathbf{q}_{1}\mathbf{,q}_{2}\right) \\
&  =\frac{\left(  2\pi\right)  ^{3/2}}{2c}\left(
\begin{array}
[c]{c}%
\frac{Q_{1}}{m_{1}}A_{j}^{ext}(\mathbf{p}_{1}-\mathbf{q}_{1})\varphi_{s_{1}%
}^{\dagger}\left(  i\left(  \mathbf{\sigma}_{1}\times\left(  \mathbf{p}%
_{1}-\mathbf{q}_{1}\right)  \right)  +\mathbf{q}_{1}+\mathbf{p}_{1}\right)
_{j}\varphi_{\sigma_{1}}\delta_{s_{2}\sigma_{2}}\delta^{3}\left(
\mathbf{p}_{2}-\mathbf{q}_{2}\right) \\
+\frac{Q_{2}}{m_{2}}A_{j}^{ext}(\mathbf{q}_{2}-\mathbf{p}_{2})\chi_{\sigma
_{2}}^{\dagger}\left(  i\left(  \mathbf{\sigma}_{2}\times\left(
\mathbf{p}_{2}-\mathbf{q}_{2}\right)  \right)  +\mathbf{q}_{2}+\mathbf{p}%
_{2}\right)  _{j}\chi_{s_{2}}\delta_{s_{1}\sigma_{1}}\delta^{3}\left(
\mathbf{p}_{1}-\mathbf{q}_{1}\right)
\end{array}
\right)  ,\nonumber
\end{align}
is the non-relativistic contribution, and%
\begin{align}
&  \mathcal{M}_{s_{1}s_{2}\sigma_{1}\sigma_{2}}^{\left(  2\right)  ext}\left(
\mathbf{p}_{1},\mathbf{p}_{2},\mathbf{q}_{1}\mathbf{,q}_{2}\right) \\
&  =\frac{\left(  2\pi\right)  ^{3/2}}{2c}\left(
\begin{array}
[c]{c}%
\frac{Q_{1}}{8m_{1}^{3}c^{2}}A_{j}^{ext}(\mathbf{p}_{1}-\mathbf{q}_{1}%
)\varphi_{s_{1}}^{\dagger}\left(
\begin{array}
[c]{c}%
\left(  \mathbf{p}_{1}^{2}-\mathbf{q}_{1}^{2}\right)  \left(  \mathbf{q}%
_{1}-i\left(  \mathbf{\sigma}_{1}\times\mathbf{q}_{1}\right)  \right) \\
-\left(  \mathbf{p}_{1}^{2}-\mathbf{q}_{1}^{2}\right)  \left(  \mathbf{p}%
_{1}+i\left(  \mathbf{\sigma}_{1}\times\mathbf{p}_{1}\right)  \right)
\end{array}
\right)  _{j}\varphi_{\sigma_{1}}\delta_{s_{2}\sigma_{2}}\delta^{3}\left(
\mathbf{p}_{2}-\mathbf{q}_{2}\right) \\
+\frac{Q_{2}}{8m_{2}^{3}c^{2}}A_{j}^{ext}(\mathbf{q}_{2}-\mathbf{p}_{2}%
)\chi_{\sigma_{2}}^{\dagger}\left(
\begin{array}
[c]{c}%
\left(  \mathbf{p}_{2}^{2}-\mathbf{q}_{2}^{2}\right)  \left(  \mathbf{q}%
_{2}-i\left(  \mathbf{\sigma}_{2}\times\mathbf{q}_{2}\right)  \right) \\
-\left(  \mathbf{p}_{2}^{2}-\mathbf{q}_{2}^{2}\right)  \left(  \mathbf{p}%
_{2}+i\left(  \mathbf{\sigma}_{2}\times\mathbf{p}_{2}\right)  \right)
\end{array}
\right)  _{j}\chi_{s_{2}}\delta_{s_{1}\sigma_{1}}\delta^{3}\left(
\mathbf{p}_{1}-\mathbf{q}_{1}\right)
\end{array}
\right)  ,\nonumber
\end{align}
(with $\varphi_{1}^{\dagger}=[1\ 0]$, $\varphi_{2}^{\dagger}=[0\ 1]$,
$\chi_{1}^{\dagger}=[0\ 1]$, $\chi_{2}^{\dagger}=-[1\ 0]$, and $j=1,2$) is the
lowest-order relativistic correction to the non-relativistic term
$\mathcal{M}^{\left(  1\right)  ext}$.

For a stationary uniform magnetic field $\mathbf{B}=B\mathbf{\hat{z}}$ the
non-zero Fourier components of the vector potential are%
\begin{equation}
A_{1}^{ext}(\mathbf{k})=\frac{\left(  2\pi\right)  ^{3/2}iB}{2}\delta\left(
k_{x}\right)  \frac{d\delta\left(  k_{y}\right)  }{dk_{y}}\delta\left(
k_{z}\right)  ,\ \ \ \ A_{2}^{ext}(\mathbf{k})=\mathbf{-}\frac{\left(
2\pi\right)  ^{3/2}iB}{2}\frac{d\delta\left(  k_{x}\right)  }{dk_{x}}%
\delta\left(  k_{y}\right)  \delta\left(  k_{z}\right)  .
\end{equation}

The trial state (11) is taken to be an eigenstate of total linear momentum
$\widehat{\mathbf{P}}$, total angular momentum squared $\widehat{\mathbf{J}%
}^{2}$, its projection $\widehat{J}_{3}$, parity $\widehat{\mathcal{P}}$, and
the Hamiltonian $\widehat{H}$, which corresponds to the hyperfine interaction
[8]. In the rest frame, where the total linear momentum vanishes, the
adjustable functions $F_{s_{1}s_{2}}(\mathbf{p}_{1}\mathbf{,p}_{2}%
)=F_{s_{1}s_{2}}(\mathbf{p}_{1})\delta\left(  \mathbf{p}_{1}+\mathbf{p}%
_{2}\right)  $ can be specified for two categories of relations among the
adjustable functions $F_{s_{1}s_{2}}(\mathbf{p})$:

\vskip0.4truecm

{\normalsize \noindent}\textit{(i) The spin-mixed (quasi-singlet and
quasi-triplet) states }

\noindent In this case we have $\ell=J$, and the general solution under the
condition of well-defined eigenvalues of $\widehat{\mathbf{P}}$,
$\widehat{\mathbf{J}}^{2}$, $\widehat{J}_{3}$, and $\widehat{\mathcal{P}}$ can
be expressed as [1], [8]
\begin{equation}
F_{s_{1}s_{2}}\left(  \mathbf{p}\right)  =C_{Jm_{J}}^{\left(  S_{1}\right)
m_{s_{1}s_{2}}}f_{J}^{\left(  S_{1}\right)  }(p)Y_{J}^{m_{s_{1}s_{2}}}\left(
\widehat{\mathbf{p}}\right)  +C_{Jm_{J}}^{\left(  S_{2}\right)  m_{s_{1}s_{2}%
}}f_{J}^{\left(  S_{2}\right)  }(p)Y_{J}^{m_{s_{1}s_{2}}}\left(
\widehat{\mathbf{p}}\right)  ,
\end{equation}
where $m_{11}=1$, $m_{12}=m_{21}=0$, $m_{22}=-1$. The $C_{Jm_{J}}^{\left(
S_{1,2}\right)  m_{s_{1}s_{2}}}=\left\langle \ell m_{\ell}Sm_{S}\mid
Jm_{J}\right\rangle $ are the Clebsch-Gordan (CG) coefficients with total spin
$S$, where $S=0$\ (with index $S_{1}$) for the singlet states and $S=1$\ (with
index $S_{2}$) for the triplet\ states respectively. Here $f_{J}^{\left(
S_{1}\right)  }(p)$ and $f_{J}^{\left(  S_{2}\right)  }(p)$ are radial
functions to be determined. They represent the contributions of spin-singlet
and spin-triplet states (the total spin $S=0,1$ is not conserved in general).

\vskip0.4truecm

{\normalsize \noindent}\textit{ (ii) The }$\ell$-\textit{mixed triplet states}

\noindent These states occur for $\ell_{1,2}=J\mp1$. Their radial
decomposition can be written as%
\begin{equation}
F_{s_{1}s_{2}}\left(  \mathbf{p}\right)  =C_{Jm_{J}}^{\left(  \ell_{1}\right)
m_{s_{1}s_{2}}}f_{\ell_{1}}(p)Y_{\ell_{1}}^{m_{s_{1}s_{2}}}\left(
\widehat{\mathbf{p}}\right)  +C_{Jm_{J}}^{\left(  \ell_{2}\right)
m_{s_{1}s_{2}}}f_{\ell_{2}}(p)Y_{\ell_{2}}^{m_{s_{1}s_{2}}}\left(
\widehat{\mathbf{p}}\right)  .
\end{equation}
Again, the $C_{Jm_{J}}^{\left(  \ell_{1,2}\right)  m_{s_{1}s_{2}}%
}=\left\langle \ell_{1,2}m_{\ell_{1,2}}Sm_{S}\mid Jm_{J}\right\rangle $ are CG
coefficients. For these states the system is characterized by $J,$ $m_{J},$
and $\mathcal{P}=(-1)^{J}$. The orbital angular momentum $\ell=\ell_{1,2}$ is
not a good quantum number. Mixing of this type occurs only for states with
principal quantum number $n\geq3$.

From the variational principle we obtain a system of coupled radial equations
expressed in matrix form as
\begin{equation}
\left(  \omega_{p}+\Omega_{p}-E\right)  \mathbb{F}\left(  p\right)
=\frac{m_{1}m_{2}}{\left(  2\pi\right)  ^{3}}\int\frac{q^{2}dq}{\sqrt
{\omega_{p}\omega_{q}\Omega_{p}\Omega_{q}}}\mathbb{K}\left(  p,q\right)
\mathbb{F}\left(  q\right)  ,
\end{equation}
where $\omega_{p}^{2}=\mathbf{p}^{2}+m_{1}^{2}$ and $\Omega_{p}^{2}%
=\mathbf{p}^{2}+m_{2}^{2}$, and $q=\left\vert \mathbf{q}\right\vert $. Here
$\mathbb{F}\left(  p\right)  $\ is the two-component matrix of radial
functions%
\begin{equation}
\mathbb{F}\left(  p\right)  =\left[
\begin{array}
[c]{c}%
f_{J}^{\left(  S_{1}\right)  }(p)\\
f_{J}^{\left(  S_{2}\right)  }(p)
\end{array}
\right]  ,\ \left[
\begin{array}
[c]{c}%
f_{\ell_{1}}(p)\\
f_{\ell_{2}}(p)
\end{array}
\right]
\end{equation}
for spin-mixed and $\ell$-mixed states respectively. The kernel of this
equation is the $2\times2$ matrix $\left[  \mathbb{K}\right]  _{ij}%
\mathbb{=}\mathcal{K}_{ij}$, which has the following form%
\begin{align}
\mathcal{K}_{ij}  &  =-i%
{\displaystyle\sum\limits_{s_{1}s_{2}\sigma_{1}\sigma_{2}}}
C_{Jm_{J}ij}^{s_{1}s_{2}\sigma_{1}\sigma_{2}}%
{\displaystyle\int}
d\widehat{\mathbf{q}}d\widehat{\mathbf{p}}~Y_{\ell_{i}}^{m_{s_{1}s_{2}}\ast
}\left(  \widehat{\mathbf{p}}\right)  Y_{\ell_{j}}^{m_{\sigma_{1}\sigma_{2}}%
}\left(  \widehat{\mathbf{q}}\right) \\
&  \times\left(  \mathcal{M}_{s_{1}s_{2}\sigma_{1}\sigma_{2}}^{ope}\left(
\mathbf{p}_{1},\mathbf{p}_{2},\mathbf{q}_{1}\mathbf{,q}_{2}\right)
+\mathcal{M}_{s_{1}s_{2}\sigma_{1}\sigma_{2}}^{\left(  1\right)  ext}\left(
\mathbf{p}_{1},\mathbf{p}_{2},\mathbf{q}_{1}\mathbf{,q}_{2}\right)
+\mathcal{M}_{s_{1}s_{2}\sigma_{1}\sigma_{2}}^{\left(  2\right)  ext}\left(
\mathbf{p}_{1},\mathbf{p}_{2},\mathbf{q}_{1}\mathbf{,q}_{2}\right)  \right)
.\nonumber
\end{align}
Here the $C_{Jm_{J}ij}^{s_{1}s_{2}\sigma_{1}\sigma_{2}}$ are related to the CG
coefficients by: $C_{Jm_{J}ij}^{s_{1}s_{2}\sigma_{1}\sigma_{2}}=C_{Jm_{J}%
}^{\left(  S_{i}\right)  m_{s_{1}s_{2}}}C_{Jm_{J}}^{\left(  S_{j}\right)
m_{s_{1}s_{2}}}$\ and $C_{Jm_{J}ij}^{s_{1}s_{2}\sigma_{1}\sigma_{2}}%
=C_{Jm_{J}}^{\left(  \ell_{i}\right)  m_{s_{1}s_{2}}}C_{Jm_{J}}^{\left(
\ell_{j}\right)  m_{\sigma_{1}\sigma_{2}}}$ for the spin- and $\ell$-mixed
states respectively. For the spin-mixed states we should take $\ell_{i}%
\equiv\ell_{j}\equiv\ell$.

The solution of equation (22) with kernel (24) including only the first two
terms with the $\mathcal{M}^{ope}$ and $\mathcal{M}^{\left(  1\right)  ext}$
matrices was discussed in [2]. This solution describes the Zeeman splitting of
the HFS energy levels in the non-relativistic limit. These energy levels are
given by formulae (1)-(4).

In order to obtain the Land\'{e} factors to order $O\left(  \alpha^{\prime
2}\right)  $ we solve the radial equation (22) with the additional term
$\mathcal{M}^{\left(  2\right)  ext}$ in the kernel (24), which is evaluated
perturbatively. The energy eigenvalues can be calculated from the matrix
equation (22) as follows:%
\begin{align}
E\int p^{2}dp\mathbb{F}^{\dagger}\left(  p\right)  \mathbb{F}\left(  p\right)
&  =\int p^{2}dp\left(  \omega_{p}+\Omega_{p}\right)  \mathbb{F}^{\dagger
}\left(  p\right)  \mathbb{F}\left(  p\right) \\
&  -\frac{m_{1}m_{2}}{\left(  2\pi\right)  ^{3}}\int_{0}^{\infty}\frac
{p^{2}dp}{\sqrt{\omega_{p}\Omega_{p}}}\int_{0}^{\infty}\frac{q^{2}dq}%
{\sqrt{\omega_{q}\Omega_{q}}}\mathbb{F}^{\dagger}\left(  p\right)
\mathbb{K}\left(  p,q\right)  \mathbb{F}\left(  q\right)  .\nonumber
\end{align}
In [2] we show that this system decouples for the spin-mixed states if the
radial functions $f_{J}^{\left(  S_{1}\right)  }$ and $f_{J}^{\left(
S_{1}\right)  }$ are taken as $f_{J}^{\left(  S_{1}\right)  }=\sqrt{\left(
1\pm\xi\right)  /2}f_{J}$ and$\ f_{J}^{\left(  S_{1}\right)  }=\mp
\sqrt{\left(  1\mp\xi\right)  /2}f_{J}$. Here $f_{J}\equiv f_{\ell}$ is a
common radial function, the upper and lower signs correspond to $sg_{q}$ and
$tr_{q}$\ states respectively. The energy corrections for $\ell$-mixed states
can also be calculated independently for $\ell=J-1$ and $\ell=J+1$ states with
corresponding radial functions $f_{\ell=J\mp1}$ (see [11]). We evaluate (25)
perturbatively using hydrogen-like radial functions (non-relativistic
Schr\"{o}dinger form $f_{\ell}=f_{n,J,m_{J}}^{Sch}\left(  p\right)  $) in
momentum space [4]).

The calculations are straightforward, and yield the relativistic corrections
to the $g$-factor for both particles of the system (the mass factors $\mu_{i}$
are defined by (5), indexes $i$ and $j$ are defined as in section 1).

\vskip0.2truecm

{\normalsize \noindent}\textit{For the triplet states} $l=J-1$,%

\begin{equation}
\bigtriangleup g_{i}^{REL}=-\frac{\mu_{j}^{2}}{2}\left(  \mu_{j}+\frac{\mu
_{i}}{J}-\frac{1}{2J+1}\right)  \left(  \frac{\alpha^{\prime}}{n}\right)
^{2}.
\end{equation}
{\normalsize \noindent}\textit{For the triplet states} $l=J+1$,%
\begin{equation}
\bigtriangleup g_{i}^{REL}=-\frac{\mu_{j}^{2}}{2}\left(  \mu_{j}-\frac{\mu
_{i}}{J+1}+\frac{1}{2J+1}\right)  \left(  \frac{\alpha^{\prime}}{n}\right)
^{2}.
\end{equation}
{\normalsize \noindent}\textit{For the spin-mixed states} $l=J$,%
\begin{equation}
\bigtriangleup g_{i}^{REL}=-\mu_{j}^{2}\left(  \mu_{j}+\mu_{i}\frac{1\mp\xi
}{4J\left(  J+1\right)  }\right)  \left(  \frac{\alpha^{\prime}}{n}\right)
^{2},
\end{equation}
where the upper and lower sign in (28) corresponds to $sg_{q}$ and $tr_{q}$
states respectively. For the spin-mixed states the parameter $\xi$\ is given
in Eq. (6).

There are several works based on the Breit equation (e.g. [12]-[15]), where
the relativistic corrections to the $g$-factor of two-fermion bound states
were considered. Some of the calculations were obtained only for small mass
ratio [12]; in others the orbital motion of the heavy particle was ignored
[13]. These calculations did not take into account the mixed nature of the
quantum states. We emphasize, that only mixed states diagonalize the
Hamiltonian of HFS [2], [8]. Our result (26)-(28) is new and overcomes the
above-mentioned shortcomings. For the relatively simple case of the ground
state $1S_{1/2}$ ($J=1$, $\mathcal{P}=-1$) our results agree with the previous
result of Hegstrom [14] and Grotch [15], namely $\bigtriangleup g_{i}%
^{REL}=-\mu_{j}^{2}\alpha^{\prime2}/3$. Note that their definition of $g_{e}$
based upon  $\Delta E_{e}=\mu_{B1}g_{e}Bm_{s}$, Eq. (12) of [14], differs by a
factor of $2$\ from our definition (1).

We provide our result (26), (28) in numerical form for the lighter particle in
atomic hydrogen (Table 1), muonium (Table 2), and muonic hydrogen (Table 3),
for which $\alpha^{\prime}=\alpha$. We consider only states with principal
quantum number $n=1,2$. For comparison, the nonrelativistic Land\'{e} factors
$g_{1}^{NR}$ ((2), (4)) with $\widetilde{g}_{s_{1}}=2$ are also included in
the tables. We used the following values for the mass ratios: $m_{e}%
/m_{p}=5.4461702\times10^{-4}$ and $m_{p}/m_{\mu}=8.8802433$. The
fine-structure constant is taken as $\alpha=7.2973523\times10^{-3}$.

\vskip0.4truecm

\begin{center}
Table 1. $g$ factor of the electron $e^{-}$ in atomic hydrogen for $n=1,2$ states.

$g_{1}^{NR}$-nonrelativistic Land\'{e} factor (2), (4), $\bigtriangleup
g_{1}^{REL}$-relativistic correction (26), (28)
\end{center}

\vskip0.2truecm

\begin{center}
$%
\begin{tabular}
[c]{||c|r|r|r|r|r||}\hline
$pe^{-}$ & $1S_{1/2\left(  J=1\right)  }$ & $2S_{1/2\left(  J=1\right)  }$ &
$2P_{1/2\left(  J=1\right)  }$ & $2P_{3/2\left(  J=1\right)  }$ &
$2P_{3/2\left(  J=2\right)  }$\\\hline
$g_{1}^{NR}$ & $1.0000000$ & $1.0000000$ & $0.3330513$ & $1.6661323$ &
$0.9997278$\\\hline
$\bigtriangleup g_{1}^{REL}$ & $-0.0000177$ & $-0.0000044$ & $-0.0000133$ &
$-0.0000133$ & $-0.0000053$\\\hline
\end{tabular}
\ $
\end{center}

\vskip0.4truecm

\begin{center}
Table 2. $g$ factor of the electron $e^{-}$ in muonium for $n=1,2$ states.

$g_{1}^{NR}$-nonrelativistic Land\'{e} factor (2), (4), $\bigtriangleup
g_{1}^{REL}$-relativistic correction (26), (28)
\end{center}

\vskip0.2truecm

\begin{center}
$%
\begin{tabular}
[c]{||c|r|r|r|r|r||}\hline
$\mu^{+}e^{-}$ & $1S_{1/2\left(  J=1\right)  }$ & $2S_{1/2\left(  J=1\right)
}$ & $2P_{1/2\left(  J=1\right)  }$ & $2P_{3/2\left(  J=1\right)  }$ &
$2P_{3/2\left(  J=2\right)  }$\\\hline
$g_{1}^{NR}$ & $1.0000000$ & $1.0000000$ & $0.3308504$ & $1.6619300$ &
$0.9975935$\\\hline
$\bigtriangleup g_{1}^{REL}$ & $-0.0000176$ & $-0.0000044$ & $-0.0000131$ &
$-0.0000131$ & $-0.0000053$\\\hline
\end{tabular}
\ $
\end{center}

\vskip0.4truecm

\begin{center}
Table 3. $g$ factor of the antimuon $\mu^{-}$ in muonic hydrogen for $n=1,2$ states.

$g_{1}^{NR}$-nonrelativistic Land\'{e} factor (2), (4), $\bigtriangleup
g_{1}^{REL}$-relativistic correction (26), (28)

\vskip0.2truecm

$%
\begin{tabular}
[c]{||c|r|r|r|r|r||}\hline
$p^{+}\mu^{-}$ & $1S_{1/2\left(  J=1\right)  }$ & $2S_{1/2\left(  J=1\right)
}$ & $2P_{1/2\left(  J=1\right)  }$ & $2P_{3/2\left(  J=1\right)  }$ &
$2P_{3/2\left(  J=2\right)  }$\\\hline
$g_{1}^{NR}$ & $1.0000000$ & $1.0000000$ & $0.2880562$ & $1.5606937$ &
$0.9496033$\\\hline
$\bigtriangleup g_{1}^{REL}$ & $-0.0000143$ & $-0.0000036$ & $-0.0000099$ &
$-0.0000098$ & $-0.0000040$\\\hline
\end{tabular}
\ \ $
\end{center}

\vskip0.4truecm

For the systems considered in Tables 1-3 the relativistic corrections for the
heavier particle $\bigtriangleup g_{2}^{REL}$ are negligible in comparison
with $\bigtriangleup g_{1}^{REL}$ due to the small mass factor $\mu_{1}$. The
relativistic corrections shown in the tables are small. They would be higher
for the corresponding states in high-$Z$ ions. For a realistic comparison of
the $g$-factor with experiment one would also need to calculate the QED
corrections up to second order in $\alpha$, that is the higher order matrix
element $\mathcal{M}^{\left(  2\right)  }$\ of Eq. (14) would have to be
included [3].

\vskip0.8truecm


\noindent\textbf{{\large 3. \textbf{Relativistic corrections to the g-factor
in the o}ne-body (Dirac) limit}}

\vskip0.4truecm

In this section we show the validity of our results for the $g$ factor in the
one-body limit. Note that the applicability of the formulae (2)-(4), (26)-(28)
is restricted by the condition $\Delta E^{ext}<<\Delta E^{HFS}$ (i.e., a weak
magnetic field $B$). To lowest order in $\alpha$\ the HFS energy splitting for
all states [8] (for $m_{2}>>m_{1}$) is given as $\Delta E^{HFS}\approx
\alpha^{\prime4}m_{1}\frac{m_{1}}{m_{2}}$. In the limit $m_{2}\rightarrow
\infty$ the HFS disappears, and in this case, the condition $\Delta
E^{ext}<<\Delta E^{HFS}$\ can not be satisfied for a nonzero magnetic field.

In this case we need to go back to the original variational equation (12) and
rewrite it in a form acceptable for the one-body limit. It is not difficult to
show that with the trial state%
\begin{equation}
\left\vert \psi_{trial}\right\rangle =\sum_{s}\int d^{3}\mathbf{p}%
F_{s}(\mathbf{p})b_{\mathbf{p}s}^{\dagger}\left\vert 0\right\rangle ,
\end{equation}
equation (12) reduces to the integral equation%
\begin{equation}
0=\sum_{s}\int d^{3}\mathbf{p}\left(  \omega_{p}-E\right)  F_{s}%
(\mathbf{p})\delta F_{s}^{\ast}(\mathbf{p})-\frac{m}{\left(  2\pi\right)
^{3}}\underset{\sigma s}{\sum}\int\frac{d^{3}\mathbf{p}d^{3}\mathbf{q}}%
{\sqrt{\omega_{p}\omega_{q}}}F_{\sigma}(\mathbf{q})\left(  -i\right)
\widetilde{\mathcal{M}}_{s\sigma}\left(  \mathbf{p,q}\right)  \delta
F_{s}^{\ast}(\mathbf{p}),
\end{equation}
where $E$ is the total one-body energy. In analogy to the two-body case, the
matrix $\widetilde{\mathcal{M}}_{s\sigma}\left(  \mathbf{p,q}\right)  $ is
made up of two parts (up to $O\left(  \alpha^{\prime4}\right)  $). The first
part corresponds to the one-photon exchange term $\mathcal{M}_{s_{1}%
s_{2}\sigma_{1}\sigma_{2}}^{ope}\left(  \mathbf{p}_{1},\mathbf{p}%
_{2},\mathbf{q}_{1}\mathbf{,q}_{2}\right)  $ taken in the limit $m_{2}%
\rightarrow\infty$. This part describes the interaction of the particle with a
static Coulomb potential $\mathcal{M}_{s_{1}s_{2}\sigma_{1}\sigma_{2}}%
^{ope}\left(  \mathbf{p}_{1},\mathbf{p}_{2},\mathbf{q}_{1}\mathbf{,q}%
_{2}\right)  \rightarrow\mathcal{M}_{s\sigma}^{Coulomb}$ [8]. The second part
represents the interaction with an external magnetic field, namely
$\mathcal{M}_{s\sigma}^{ext}\left(  \mathbf{p,q}\right)  =\mathcal{M}%
_{s\sigma}^{\left(  1\right)  ext}\left(  \mathbf{p,q}\right)  +\mathcal{M}%
_{s\sigma}^{\left(  2\right)  ext}\left(  \mathbf{p,q}\right)  $, which can be
obtained from (17)-(18)%
\begin{equation}
\mathcal{M}_{s\sigma}^{\left(  1\right)  ext}\left(  \mathbf{p,q}\right)
=\frac{\left(  2\pi\right)  ^{3/2}Q}{2mc}A_{j}^{ext}(\mathbf{p}-\mathbf{q}%
)\varphi_{s}^{\dagger}\left(  i\left(  \mathbf{\sigma}\times\left(
\mathbf{p}-\mathbf{q}\right)  \right)  +\mathbf{q}+\mathbf{p}\right)
_{j}\varphi_{\sigma}%
\end{equation}%
\begin{equation}
\mathcal{M}_{s\sigma}^{\left(  2\right)  ext}\left(  \mathbf{p,q}\right)
=\frac{\left(  2\pi\right)  ^{3/2}Q}{16m^{3}c^{3}}A_{j}^{ext}(\mathbf{p}%
-\mathbf{q})\varphi_{s}^{\dagger}\left(  \left(  \mathbf{p}^{2}-\mathbf{q}%
^{2}\right)  \left(  \mathbf{q}-i\left(  \mathbf{\sigma}\times\mathbf{q}%
\right)  \right)  -\left(  \mathbf{p}^{2}-\mathbf{q}^{2}\right)  \left(
\mathbf{p}+i\left(  \mathbf{\sigma}\times\mathbf{p}\right)  \right)  \right)
_{j}\varphi_{\sigma}%
\end{equation}
Further calculations require a classification of the states. The trial state
(29) is taken to be an eigenstate of total angular momentum squared
$\widehat{\mathbf{j}}^{2}$ and its projection $\widehat{j}_{3}$. These
conditions can be satisfied if the adjustable two-component functions
$F_{s}(\mathbf{p})$\ are taken in the following form:

\vskip0.2truecm

{\normalsize \noindent}\textit{For states} $\ell=j-1/2$%
\begin{equation}
F_{1}\left(  \mathbf{p}\right)  =f_{j-\frac{1}{2}}\left(  p\right)  C_{jm_{j}%
}^{\ell m_{\ell}\frac{1}{2}\frac{1}{2}}Y_{j-1/2}^{m_{j}-1/2}\left(
\widehat{\mathbf{p}}\right)  ,\ \ \ F_{2}\left(  \mathbf{p}\right)
=f_{j-\frac{1}{2}}\left(  p\right)  C_{jm_{j}}^{\ell m_{\ell}\frac{1}{2}%
-\frac{1}{2}}Y_{j-1/2}^{m_{j}+1/2}\left(  \widehat{\mathbf{p}}\right)
\end{equation}
{\normalsize \noindent}\textit{For states} $\ell=j+1/2$%
\begin{equation}
F_{1}\left(  \mathbf{p}\right)  =f_{j+\frac{1}{2}}\left(  p\right)  C_{jm_{j}%
}^{\ell m_{\ell}\frac{1}{2}\frac{1}{2}}Y_{j+1/2}^{m_{j}-1/2}\left(
\widehat{\mathbf{p}}\right)  ,\ \ \ F_{2}\left(  \mathbf{p}\right)
=f_{j+\frac{1}{2}}\left(  p\right)  C_{jm_{j}}^{\ell m_{\ell}\frac{1}{2}%
-\frac{1}{2}}Y_{j+1/2}^{m_{j}+1/2}\left(  \widehat{\mathbf{p}}\right)  ,
\end{equation}
where $C_{jm_{j}}^{\ell m_{\ell}sm_{s}}=\left\langle \ell m_{\ell}sm_{s}\mid
jm_{j}\right\rangle $ are the CG coefficients for $s=1/2$, $m_{s}=\pm1/2$.

After substitution of these formulae into (30) and completion of the
variational procedure we obtain the radial equation%
\begin{equation}
\left(  \omega_{p}-E\right)  f_{\ell}\left(  p\right)  =\frac{m}{\left(
2\pi\right)  ^{3}}\int\frac{q^{2}dq}{\sqrt{\omega_{p}\omega_{q}}}%
\mathcal{K}_{\ell}\left(  p,q\right)  f_{\ell}\left(  p\right)  ,
\end{equation}
where the kernel $\mathcal{K}_{\ell}\left(  p,q\right)  $ is expressed through
the matrix $\mathcal{M}$\ and coefficients $C_{jm_{j}}^{\sigma^{\prime}\sigma
}=C_{jm_{j}}^{\ell m_{\ell}sm_{\sigma^{\prime}}}C_{jm_{j}}^{\ell m_{\ell
}sm_{\sigma}}$, namely%
\begin{equation}
\mathcal{K}_{\ell}\left(  p,q\right)  =-i%
{\displaystyle\sum\limits_{\sigma^{\prime}\sigma}}
C_{jm_{j}}^{\sigma^{\prime}\sigma}%
{\displaystyle\int}
d\widehat{\mathbf{q}}d\widehat{\mathbf{p}}~\left(  \mathcal{M}_{\sigma
^{\prime}\sigma}^{Coulomb}+\mathcal{M}_{\sigma^{\prime}\sigma}^{\left(
1\right)  ext}\left(  \mathbf{p,q}\right)  +\mathcal{M}_{\sigma^{\prime}%
\sigma}^{\left(  2\right)  ext}\left(  \mathbf{p,q}\right)  \right)  Y_{\ell
}^{m_{\sigma^{\prime}}\ast}\left(  \widehat{\mathbf{p}}\right)  Y_{\ell
}^{m_{\sigma}}\left(  \widehat{\mathbf{q}}\right)  ,
\end{equation}
with $\sigma^{\prime}=1,2$, $\sigma=1,2$, and $m_{1,2}=m_{j}\mp1/2$.

In the absence of an external magnetic field, equation\ (35) represents the
relativistic radial equation in momentum space for a bound one-body system. As
was shown in [8], the solution of the two-body equation (22) reduces to the
solution of the one-body equation (35).

We now evaluate the contribution of the next two terms $\mathcal{M}%
_{\sigma^{\prime}\sigma}^{\left(  1\right)  ext}\left(  \mathbf{p,q}\right)  $
and $\mathcal{M}_{\sigma^{\prime}\sigma}^{\left(  2\right)  ext}\left(
\mathbf{p,q}\right)  $ of equations (31) and (32). The relevant energy $\Delta
E_{j,m_{j},\ell}^{ext}$ can be calculated perturbatively like in the two-body
case (cf. Eq. (25)). We obtain%
\begin{equation}
\Delta E_{j,m_{j},\ell}^{ext}=\mu_{B}g_{D}Bm_{j},
\end{equation}
where%
\begin{equation}
g_{D}=g_{L}\left(  \ell,j\right)  +\bigtriangleup g_{D}^{REL}\left(
n,j\right)  .
\end{equation}
Here $g_{L}\left(  \ell\right)  $ is the usual result for the anomalous Zeeman
effect, being the Land\'{e} $g$\ factor, whose value is ([4]):%
\begin{equation}
g_{L}\left(  \ell,j\right)  =\frac{2j+1}{2\ell+1}.
\end{equation}
The second term is the relativistic correction to the $g_{D}$\ factor%
\begin{equation}
\bigtriangleup g_{D}^{REL}\left(  n,j\right)  =-\frac{\left(  2j+1\right)
^{2}}{8j\left(  j+1\right)  }\left(  \frac{\alpha^{\prime}}{n}\right)  ^{2}.
\end{equation}
Formulae (39) and (40) coincide with the result of the expansion (up to
$O\left(  \alpha^{\prime2}\right)  $) of the general formula for the
$g$-factor obtained by Margenau [16]\ on the basis of the Dirac equation with
$\alpha^{\prime}= Z \alpha$.

\newpage

\vskip0.8truecm

\noindent\textbf{{\large 4. Concluding remarks}}

\vskip0.4truecm

We considered the relativistic theory of the Zeeman splitting and the
$g$-factor of the hyperfine structure in the two-fermion system on the basis
of variational relativistic equations derived from quantum electrodynamics
[1]. Relativistic corrections to the $g$-factor beyond the non-relativistic
formulae (2)-(4) (from Ref. [2]) were calculated up to order $O\left(
\alpha^{\prime2}\right)  $ for all quantum states, and are given in
Eqs.~(26)-(28). The $g$-factor\ corrections take into account the mixed nature
of the states (spin-mixing and $\ell$-mixing), and the orbital motion of the
heavy particle. They are obtained for arbitrary mass ratio and are symmetrical
with respect to the masses of the constituent particles. For the ground state
our result reduces to the well-known formula obtained some time ago by
Hegstrom [14] and Grotch [15]. We show that in the one-body case the solution
of the variational relativistic equation reproduces the result for the
$g$-factor obtained from the Dirac equation.

In our calculations we assumed that the trial state (11) is an eigenstate of
the total linear momentum operator $\widehat{P}$. However this is only an
approximation, because one can show that the operator $\widehat{P}$\ does not
commute with the HFS Hamiltonian. To fix this problem we need to modify the
trial state, or use an appropriate unitary transformation for the HFS
Hamiltonian. The latter approach was discussed in the literature [12], [13],
[17]. An analysis shows, that in our case the unitary transformation leads to
the appearance of additional terms in the invariant $\mathcal{M}$-matrix. This
is a technically difficult problem which we postpone for the future.

We note that, in contrast to the Breit approach, the anomalous magnetic moment
is not introduced from another calculation. As discussed in section 2 (below
equation (13)) all QED effects are contained in the $\mathcal{M}$-matrix. The
anomalous magnetic moment will appear naturally in our calculations if we
include the next term $\mathcal{M}^{\left(  2\right)  }$ of the expansion of
$\mathcal{M}$- matrix in (13).

Concerning comparison with experiment we note that measurements so far appear
to be restricted to states where the spin structure of the heavier particle
can be ignored. This includes data for the $1S_{1/2}$\ state in ions [18],
[19]. The present work will be of practical importance when these measurements
will be extended to all $n=2$\ (or higher-$n$) levels.

\vskip0.8truecm

{\normalsize \noindent{\textbf{{\large Acknowledgment}}} }

{\normalsize \vskip0.4truecm }

AGT and MH acknowledge the financial support of the Natural Sciences and
Engineering Research Council of Canada for this work.

\newpage

{\normalsize \vskip0.8truecm }

{\normalsize \noindent{\textbf{{\large References}}} }

{\normalsize \vskip0.4truecm }

{\normalsize \enumerate}

1. A. G. Terekidi, J. W. Darewych, M. Horbatsch, Can. J. Phys. \textbf{85}, 813-836.(2007).

2. A. G. Terekidi, J. W. Darewych, M. Horbatsch, Phys. Rev. A \textbf{75},
043401 (2007).

3. V. W. Hughes and G. zu Putlitz, in \textit{Quantum Electrodynamics}, edited
by T. Kinoshita (World Scientific, Singapore, 1990), p. 822.

4. H. A. Bethe and E. E. Salpeter, \textit{Quantum Mechanics of One- and
Two-Electron Atoms }(Springer, 1957).

5. M. Mizushima, \textit{Quantum Mechanics of Atomic Spectra and Atomic
Structure} (W. A. Benjamin, 1970). p.331.

6. G. K. Woodgate, \textit{Elementary atomic structure} (Clarendon Press,
Oxford, 1980).

7. J. Schwinger, Phys. Rev. \textbf{73}, 416 (1948).

8. A. G. Terekidi, J. W. Darewych, Journal of Mathematical Physics
\textbf{46}, 032302 (2005).

9. J. W. Darewych, Annales Fond. L. de Broglie (Paris) \textbf{23}, 15 (1998).

10. J. W. Darewych, in \textit{Causality and Locality in Modern Physics}, G
Hunter et al. (eds.), p. 333, (Kluwer, 1998).

11. A. G. Terekidi, J. W. Darewych, Journal of Mathematical Physics
\textbf{45}, 1474 (2004).

12. H. Grotch, R. Kashuba, Phys. Rev. A, \textbf{7}, 78-84 (1973)

13. J. M. Anthony, K. J. Sebastian, Phys. Rev. A, \textbf{49}, 192-206 (1994)

14. R. A. Hegstrom, Phys. Rev. \textbf{184}, 1, p. 17-22 (1969).

15. H. Grotch, Phys. Rev. Let. \textbf{24}, 2, p.39-42 (1970).

16. H. Margenau, Phys. Rev. \textbf{57}, p. 387-386 (1940).

17. H Grotch, Roger A. Hegstrom, Phys. Rev. A, \textbf{4}, 59-69 (1971).

18. J. L. Verd\'{u}, S. Djeki\'{c}, S. Stahl, T. Valenzuela, M. Vogel, G.
Werth, T. Beier, H. -J. Kluge, and W. Quint, Phys. Rev. Lett. \textbf{92},
093002 (2004).

19. D. L. Moskovkin, V. M. Shabaev, Phys. Rev. A, \textbf{73},.052506 (2006).

\end{document}